\documentclass[twoside,web]{ieeecolor}
\usepackage{generic}
\usepackage{cite}
\usepackage{amsmath,amssymb,amsfonts}
\usepackage{algorithmic}
\usepackage{graphicx}
\usepackage{siunitx}
\usepackage{textcomp}
\usepackage{bm}
\makeatletter
\let\NAT@parse\undefined
\makeatother
\usepackage[colorlinks=true]{hyperref}
\hypersetup{
	citecolor=blue,
	linkcolor=blue,
	urlcolor=blue}
\def\BibTeX{{\rm B\kern-.05em{\sc i\kern-.025em b}\kern-.08em
		T\kern-.1667em\lower.7ex\hbox{E}\kern-.125emX}}
\usepackage{orcidlink}
\begin{document}
\title{Theoretical Limit of MOSFET Subthreshold Swing at Sub-Kelvin Temperatures}
\author{Arnout Beckers \orcidlink{0000-0003-3663-0824} 
\thanks{This work is supported by the Chips JU project ARCTIC (Project \#101139908). The project is supported by the Chips Joint Undertaking and its members (including top-up funding by Belgium, Austria, Germany, Estonia, Finland, France, Ireland, The Netherlands and Sweden). ARCTIC gratefully acknowledges the support of the Canadian and the Swiss federal governments.}
\thanks{A. Beckers is with imec, Kapeldreef 75, 3001 Leuven, Belgium.}}

\maketitle

\begin{abstract}
Fully conductive band tails cause the subthreshold swing to saturate at temperatures above 1\,K. However, recent measurements indicate that below 1\,K, the subthreshold swing in certain MOSFET structures resumes a linear scaling with temperature. Following this ultra-steep behavior, a new type of plateau has been measured below 1\,K. In this letter, we show that hybrid band tails, with both traps and mobile states, explain this new plateau. Furthermore, hybrid band tails explain various non-saturating behaviors above 1\,K. Remarkably, for entirely non-conductive band tails, the simulations and theory predict a third type of plateau below 10\,mK. We hypothesize that this represents the lower bound of subthreshold swing at sub-Kelvin temperatures, which is a testable prediction from the theory. 
\end{abstract}


\section{Introduction}
\begin{figure}
	\centering
	\includegraphics[width=0.35\textwidth]{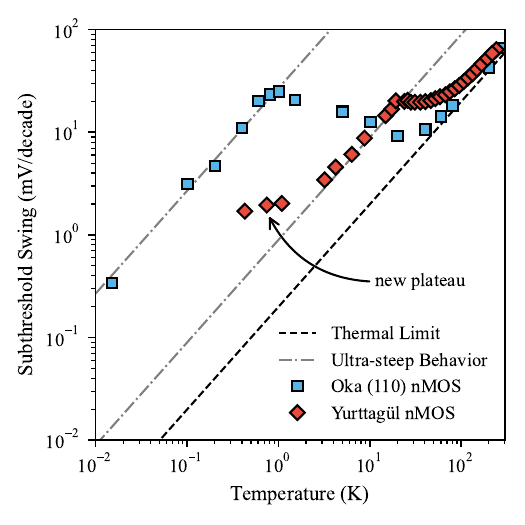}
	\vspace{-0.4cm}
	\caption{Experimental subthreshold swing from \cite{oka,yurtta} in log-log scale.}
	\label{fig:1}
\end{figure}

It is a long-standing research question how steeply MOSFETs can theoretically switch at low temperatures \cite{tewksbury}. The inverse subthreshold slope, or subthreshold swing, $SS=\mathrm{d}V_{\mathrm{GB}}/\mathrm{d}\log_{10}(I_{\mathrm{DS}})$, is expected to decrease linearly with temperature, and then saturate to a plateau  below $T_\mathrm{c}\approx \SI{50}{\kelvin}$ due to a band tail with decay parameter $k_\mathrm{B}T_\mathrm{c}$ \cite{bohus,edl,ghibaudo,tnano,ghibaudo2,hafez,jock}. However, Oka \emph{et al.} reported $SS$ in a large MOSFET to scale again linearly with $T$ below 1\,K (see Fig.\ref{fig:1}) \cite{oka}. Furthermore, Yurttag\"{u}l \emph{et al.} measured a new plateau below 1\,K \cite{yurtta}, which cannot be explained by the saturation model from \cite{bohus,edl,ghibaudo,tnano} assuming a conductive band tail, nor by the thermal ultra-steep model from Oka \emph{et al.} assuming a non-conductive band tail \cite{oka}. In this letter, three main improvements over these works are presented: (i) a mobility edge inside the band tail dividing traps and mobile states, i.e., a \textquotedblleft hybrid\textquotedblright \, band tail, (ii) an accurate Fermi level from numerical simulations that transcends into the band, and (iii) a smooth transition between tail and band.
\section{\label{sec:model}Subthreshold Swing Model}The subthreshold swing temperature dependence, $SS(T)$, 
is determined by first solving the coupled equations
\begin{align}
	I_{\mathrm{DS}} &= \frac{W}{L} \cdot \mu(T)\cdot q\cdot n_{\mathrm{mobile}}(E_{\mathrm{F}},T,T_\mathrm{c},E_{\mu})\cdot V_{\mathrm{DS}} \label{eq:ids} \\
	V_{\mathrm{GB}}^\prime &= \frac{E_{\mathrm{F}} - E_{\mathrm{c}}}{q} + \frac{q\cdot n_{\mathrm{total}}(E_{\mathrm{F}},T,T_\mathrm{c})}{C_{\mathrm{ox}}} \label{eq:vgb}
\end{align}
where $E_{\mu}$ is the mobility edge, $q$ is the elementary charge, and other symbols retain their conventional meanings. The electron mobility, $\mu(T)$, is evaluated at constant $I_{\mathrm{DS}}$ at which $SS(T)$ is extracted at small $V_{\mathrm{DS}}$ in the linear regime. Interface-trap density constant over energy, gate-metal work function, and fixed charges due to doping are implicitly assumed in $V_{\mathrm{GB}}^\prime$. The standard slope factor for $E_\mathrm{F}\!\ll\! E_\mathrm{c}$, given by $m = 1 + (C_{\mathrm{depl}} + C_{\mathrm{it}})/C_{\mathrm{ox}}$, is included in the final expressions where needed. Here $C_{\mathrm{it}}$ accounts only for the uniform trap density over energy, excluding traps in the exponential tail.
\begin{figure}[t]
	\centering
	\includegraphics[width=0.28\textwidth]{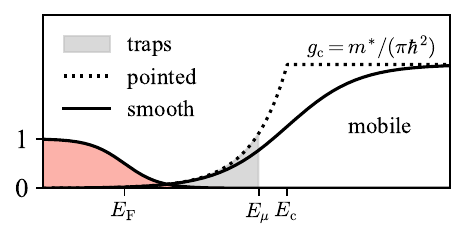}
	\vspace{-0.4cm}
	\caption{Fermi-Dirac function (left) and pointed versus smooth density-of-states models (right). The mobility edge ($E_\mathrm{\mu}$) separates traps and mobile states. Different band-tail types can be studied by moving $E_\mathrm{\mu}$ over the band tail: (1) $E_\mathrm{\mu} \ll E_\mathrm{c}$: fully conductive band tail, (2) $E_\mathrm{\mu} < E_\mathrm{c}$: hybrid band tail, and (3) $E_\mathrm{\mu}=E_\mathrm{c}$: non-conductive band tail.} 
	\label{fig:2}
\end{figure}\begin{figure*}[t]
	\includegraphics[width=\textwidth]{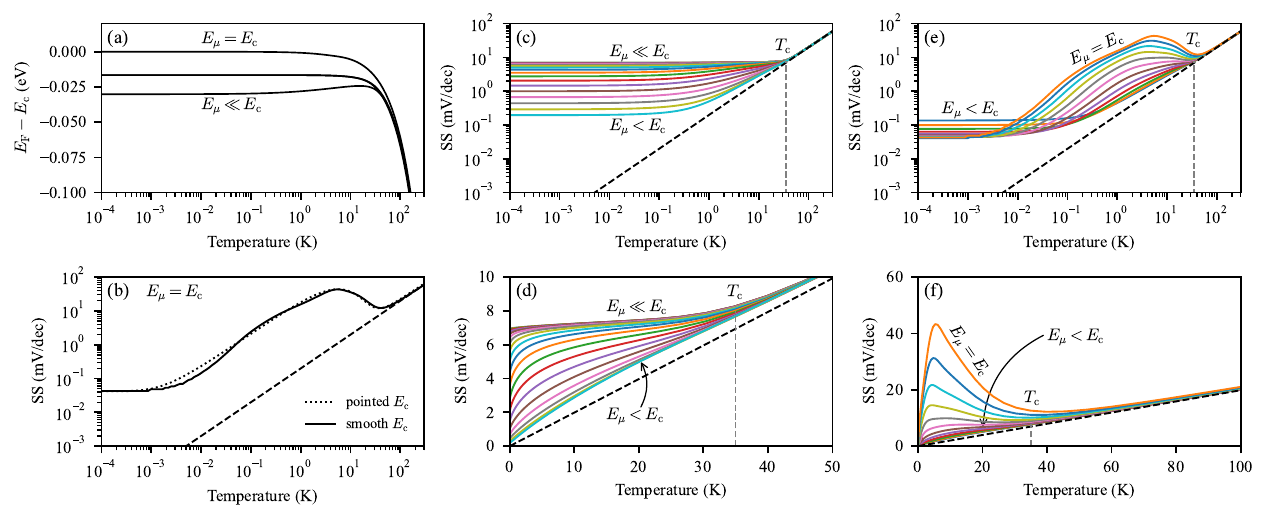}
	\vspace{-0.8cm}
	\caption{Numerical simulations show the impact of moving $E_\mathrm{\mu}$ over the band tail. (a) For conductive tails ($E_\mathrm{\mu}\!\ll\! E_\mathrm{c}$) $E_\mathrm{F}$ settles below $E_\mathrm{c}$. For non-conductive band tails ($E_\mathrm{\mu}\!=\!E_\mathrm{c}$) $E_{\mathrm{F}}$ approaches $E_\mathrm{c}$ and transcends it ($E_{\mathrm{F}}=E_\mathrm{c}+35\cdot k_\mathrm{B}T$ at \SI{0.1}{\milli\kelvin}), (b) A smooth band edge makes the ultra-steep behavior less straight, (c) Moving $E_\mathrm{\mu}$ toward $E_\mathrm{c}$ causes a lower plateau, (d) In linear $T$-scale, a sudden drop of the standard plateau can be seen around a few Kelvins, and an almost straight line close to the thermal limit for $E_{\mu}<E_\mathrm{c}$ ($E_\mathrm{\mu}=E_\mathrm{c}-5\cdot k_\mathrm{B}T_\mathrm{c}$), (e) Further moving $E_\mathrm{\mu}$ closer to $E_\mathrm{c}$ causes the large bump due trap filling, (f) In linear $T$-scale, $SS$ transitions to a linear trend with a high slope ($m^{\prime\prime}$). Parameters: $T_\mathrm{c}=\SI{35}{\kelvin}$, $\mu_0=\SI{2000}{\centi\meter\squared\per\volt\per\second}$, \SI{5}{\nano\meter} SiO$_2$, $I_{\mathrm{DS}}=\SI{e-10}{\ampere}$. The colors in (c) and (d) match each other, as well as in (e) and (f).} 
	\label{fig:3}
\end{figure*}

The mobile and total electron densities can be computed using pointed or smooth density-of-states models, as shown in Fig.~\ref{fig:2}. In the pointed model, an exponential function with decay parameter $k_\mathrm{B}T_\mathrm{c}$ is connected to the rectangular band, which allows to express the electron density in the band tail as a Gauss hypergeometric function\footnote{For details on recasting a Fermi-Dirac integral with exponential density-of-states into a Gauss hypergeometric function, see Section III in~\cite{gen}.} $H_{2F1}$, yielding  $n_\mathrm{total}=g_\mathrm{c}\cdot k_\mathrm{B}T_\mathrm{c}\cdot H_{2F1}\left(1, \theta, \theta+1, z\right)+g_\mathrm{c}\cdot k_\mathrm{B}T\cdot \ln\left(1+e^{u}\right)$ and $n_\mathrm{mobile}=n_\mathrm{total}-g_\mathrm{c}\cdot k_\mathrm{B}T_\mathrm{c}\cdot \exp(w\cdot\theta)\cdot H_{2F1}\left(1,\theta,\theta+1,z\cdot e^w\right)$, where $z=-\exp(-u)$,  $u=(E_\mathrm{F}-E_\mathrm{c})/(k_\mathrm{B}T)$, $w=(E_\mathrm{\mu}-E_\mathrm{c})/(k_\mathrm{B}T)$, and $\theta=T/T_\mathrm{c}$. In the smooth density-of-states model, the band tail and band are described by a sigmoid function with decay parameter $k_{\mathrm{B}} T_{\mathrm{c}}$, which must be numerically integrated over the Fermi–Dirac function.

\section{Numerical Results for Fermi level and Subthreshold Swing}
Fig.\ref{fig:3} shows the simulated $E_\mathrm{F}(T)$ and $SS(T)$ while sweeping $E_\mu$ from far below $E_\mathrm{c}$ (tail is entirely conductive) to $E_\mathrm{c}$ (tail is non-conductive). In Fig.\ref{fig:3}(a), $\vert E_\mathrm{F}-E_\mathrm{c}\vert$ reduces to satisfy the constant current condition. The landing position of $E_\mathrm{F}$ below \SI{1}{\kelvin} depends on $E_\mu$. Fig.\ref{fig:3}(b) shows the impact of smoothing the density-of-states around the band edge causing the ultra-steep behavior to be less straight. In Fig.\ref{fig:3}(c), it can be seen that $E_\mathrm{\mu}<E_\mathrm{c}$ can explain the new plateau. Furthermore, Fig.\ref{fig:3}(d) shows that $SS$ above 1\,K suddenly drops below the plateau at a few degrees Kelvin, or scales almost linearly close to the thermal limit. Fig.\ref{fig:3}(e) shows that the size of the bump due to trap filling can also be explained by $E_\mathrm{\mu}$, not only by crystal orientation as in \cite{oka}. Fig.\ref{fig:3}(f) shows the increase and sharp drop of $SS$ for hybrid and non-conductive band tails.
\begin{figure}[t]
	\centering
	\includegraphics[width=0.45\textwidth]{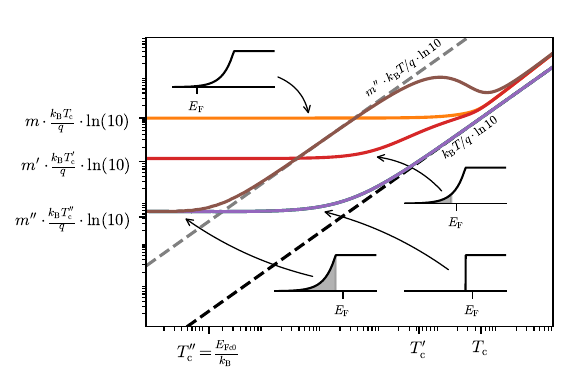}
	\vspace{-0.5cm}
	\caption{$SS(T)$ saturates if $E_\mathrm{F}$ lands in a conductive section of the density of states (band tail or band). Fully conductive and no band tails therefore produce two phases (Boltzmann directly followed by a plateau because no traps are being filled). Hybrid and non-conductive tails produce four phases (Boltzmann, bump, linear scaling, and a plateau).}
	\label{fig:4}
\end{figure}

Fig.\ref{fig:4} shows the three different types of plateaus. We summarize the main points for each type of band tail:
\begin{itemize}
	\item \textbf{conductive band tail ($\bm{E_\mathrm{\mu}\!\ll\!E_\mathrm{c}}$)}: $E_\mathrm{F}$ finds mobile states at the bottom of the tail (top left inset), and settles there, explaining why the standard plateau, $m\cdot k_\mathrm{B}T_\mathrm{c}/q\cdot\ln(10)$, starts early (at $T>$\,1\,K) and continues below 1\,K. 
	
	\item \textbf{hybrid band tail ($\bm{E_\mathrm{\mu}\!<\!E_\mathrm{c}}$)}: $SS$ saturates at lower temperature ($T_\mathrm{c}^\prime$) because $E_\mathrm{F}$ first needs to reach the conductive section of the tail and cross traps (center right inset). A hybrid band tail has only a fraction of the traps of a non-conductive tail, which explains the modest bump in Yurttag\"{u}l's data in Fig.\ref{fig:1} compared to Oka's. Furthermore, less mobile states starting at higher energies explains why Yurttag\"{u}l's data has a lower plateau than the standard plateau. We derive the formula for this new plateau, $m^\prime \cdot k_\mathrm{B}T_\mathrm{c}^\prime/q\cdot\ln(10)$, in Section \ref{sec:plateau}. 
	
	\item \textbf{non-conductive band tail ($\bm{E_\mathrm{\mu}\!=\!E_\mathrm{c}}$)}: Oka \emph{et al.} derived the ultra-steep behavior $m^{\prime\prime}\cdot k_\mathrm{B}T/q\cdot\ln(10)$, assuming $E_\mathrm{F}$ at $E_{\mathrm{c}}$ in the 0-K limit \cite{oka}. However, our numerical simulations in Fig.\ref{fig:3}(a) indicate that the limit of $E_{\mathrm{F}}$ is not $E_\mathrm{c}$, as in Oka's model, but lies several $k_\mathrm{B}T$ beyond $E_{\mathrm{c}}$ at $T<$\,10\,mK. As will be derived in Section \ref{sec:plateau}, an $E_\mathrm{F}$ inside the mobile band (bottom left inset) makes $SS$ saturate to $m^{\prime\prime}\cdot k_\mathrm{B}T_{c}^{\prime\prime}/q\cdot \ln(10)$ where $T_{c}^{\prime\prime}=E_\mathrm{Fc0}/k_\mathrm{B}$. 
	\item \textbf{no band tail}: (bottom right inset) $SS$ follows the thermal limit and then the same plateau proportional to $T_\mathrm{c}^{\prime\prime}$. In a pristine channel with no disorder, $SS$ saturates also. 
\end{itemize}

\begin{figure}[t]
	\centering
	\includegraphics[width=0.4\textwidth]{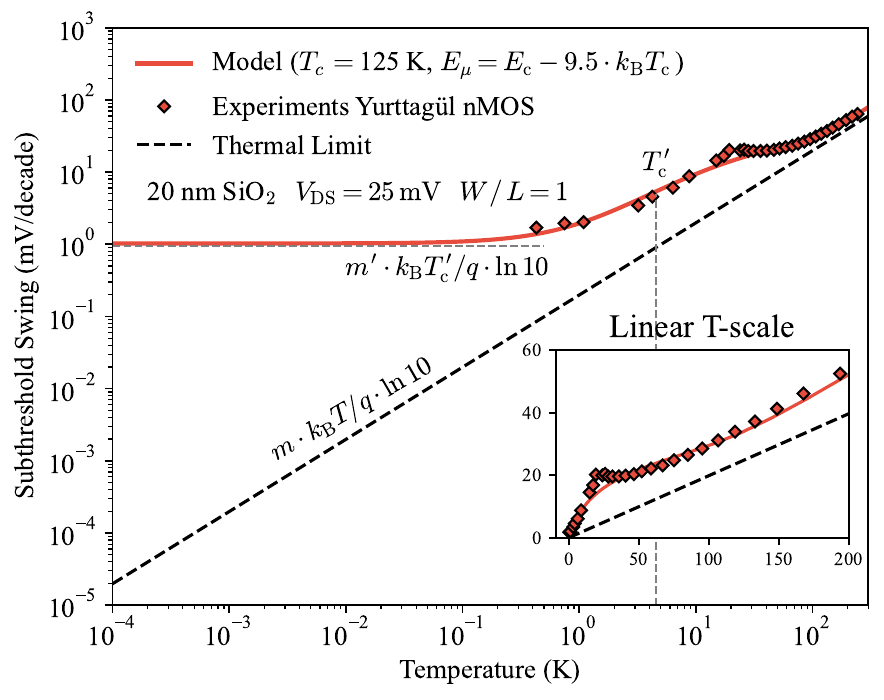}
	\caption{Fit of Yurttag\"{u}l's data, using a smooth band edge, a hybrid band tail, $I_\mathrm{DS}=\SI{e-10}{\ampere}$, and other parameters as in the figure. The bump is smaller because less traps are being filled than in Oka's data.}
	\label{fig:yurtta}
\end{figure}

\section{\label{sec:plateau}Derivation of Two New Plateaus} 
For a hybrid band tail, $SS$ saturates if $E_\mathrm{\mu}\! <\! E_\mathrm{F}\!<\! E_\mathrm{c}$. In the 0-K limit (step Fermi-Dirac function), the electron densities in (\ref{eq:ids})-(\ref{eq:vgb}) are $n_{\mathrm{total}}=g_\mathrm{c}\cdot k_\mathrm{B}T_\mathrm{c}\cdot\exp\left(E_\mathrm{Fc}/k_\mathrm{B}T_\mathrm{c}\right)$ and $n_\mathrm{mobile}=n_\mathrm{total}-g_\mathrm{c}\cdot k_\mathrm{B}T_\mathrm{c}\cdot\exp\left(E_\mathrm{\mu c}/k_\mathrm{B}T_\mathrm{c}\right)$, where $E_\mathrm{Fc}=E_\mathrm{F}-E_\mathrm{c}$ and $E_\mathrm{\mu c}=E_\mathrm{\mu}-E_\mathrm{c}$. Since $\mu$ in (\ref{eq:ids}) is assumed independent of $E_{\mathrm{Fc}}$, the expression for $SS$ can be written as  
\begin{equation}
	SS=\frac{\mathrm{d}V_\mathrm{GB}^\prime}{\mathrm{d}E_\mathrm{Fc}}\cdot n_\mathrm{mobile}\cdot\left(\frac{\mathrm{d}n_\mathrm{mobile}}{\mathrm{d}E_{\mathrm{Fc}}}\right)^{-1}\cdot \ln(10)
	\label{eq:ss}
\end{equation}
Differentiating $V_{\mathrm{GB}}^\prime$ from (\ref{eq:vgb}) and $n_\mathrm{mobile}$ to $E_{\mathrm{Fc}}$, combining that in (\ref{eq:ss}), and including the implicit $m$, gives
\begin{equation}
	SS=m^\prime\cdot \frac{k_\mathrm{B}T_\mathrm{c}^\prime}{q}\cdot \ln(10),
	\label{eq:newplateau}
\end{equation}
where $m^\prime = m+q^2\cdot g_\mathrm{c}\cdot e^{x}/C_\mathrm{ox}$ and $T_{c}^\prime=T_\mathrm{c}\cdot\left(1-e^{y-x}\right)$
with $x=E_\mathrm{Fc0}/(k_\mathrm{B}T_\mathrm{c})$ and $y=E_\mathrm{\mu c}/(k_\mathrm{B}T_\mathrm{c})$. $E_{\mathrm{Fc0}}$ is the saturating value of $E_\mathrm{F}-E_\mathrm{c}$ in the 0-K limit. If $E_\mathrm{\mu}\ll E_\mathrm{c}$ ($e^y\rightarrow 0$), (\ref{eq:newplateau}) falls back to the standard plateau for a conductive tail, consistent with the simulations in Fig.\ref{fig:3}(c). $T_\mathrm{c}^\prime$ is below $T_\mathrm{c}$ because $E_\mathrm{F}>E_\mathrm{\mu}$ (i.e, $y-x<0$). 

Fig.\ref{fig:yurtta} validates the hybrid band-tail model with the measurements from \cite{yurtta} in logarithmic and linear (inset) scales. The correct experimental values for SiO$_2$ thickness, $V_{\mathrm{DS}}$, $W/L$, and $I_{\mathrm{DS}}$ are set in the model. We assume a monotonic increase and saturation of $\mu(T)$, from \SI{200}{\centi\meter\squared\per\volt\per\second} at \SI{300}{\kelvin} to \SI{2000}{\centi\meter\squared\per\volt\per\second} at \SI{0.1}{\milli\kelvin}, consistent with experimental trends down to $\sim$\SI{400}{\milli\kelvin} \cite{yurtta}, and extrapolated beyond. This extrapolation seems justified by the low doping levels in the devices, which suppress Coulomb scattering and allow $\mu(T)$ to continue saturating. Using $E_{\mathrm{Fc0}}=\SI{-0.10193}{\electronvolt}$ (from numerical simulations) in (\ref{eq:newplateau}) ($m\!=\!1$, $m^*\!=\!1.08\cdot m_\mathrm{e}$, $g_\mathrm{c}\!=\!m^*/(\pi\hbar^2)=\SI{4.51e14}{\per\centi\meter\squared\per\electronvolt}$), we obtain $SS=\SI{0.94}{\milli\volt}/\mathrm{dec}$ and $T_{c}^\prime=\SI{4.6}{\kelvin}$, annotated in Fig.\ref{fig:yurtta}. \begin{figure}[t]
	\centering
	\includegraphics[width=0.4\textwidth]{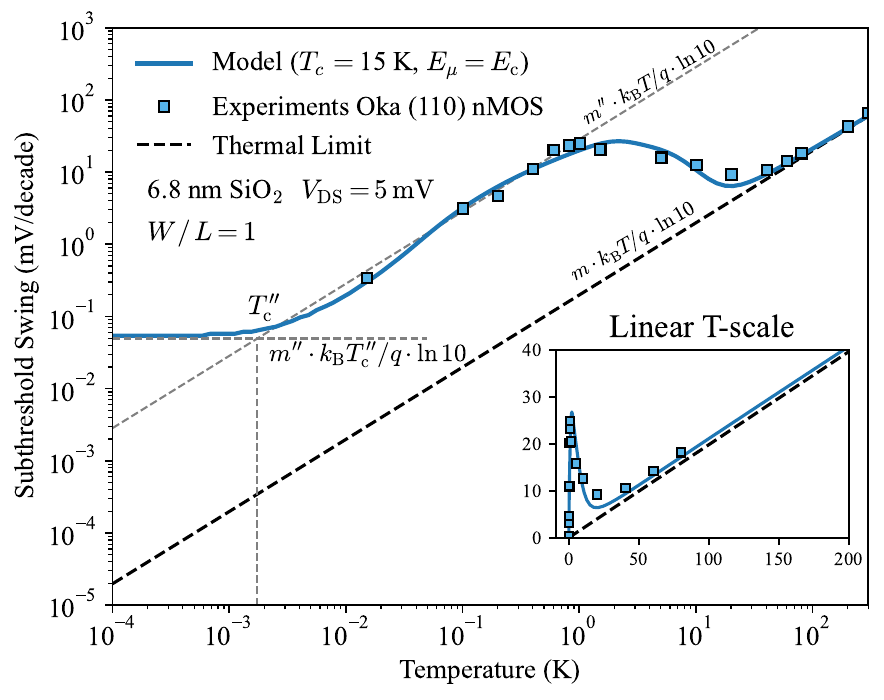}
	\caption{Fit of Oka's data using a smooth band edge, a non-conductive band tail, $I_\mathrm{DS}\!=\!\SI{e-10}{\ampere}$, $\mu_0\!=\!\SI{2000}{\centi\meter\squared\per\volt\per\second}$, and other parameters as in the figure. The model predicts a saturation below $T_\mathrm{c}^{\prime\prime}$.}
	\label{fig:oka}
\end{figure}

In Fig.\,\ref{fig:oka}, the non-conductive band-tail model agrees with the mK measurements from \cite{oka} down to $\approx$\,\SI{15}{\milli\kelvin}. Below $\approx\,\SI{15}{\milli\kelvin}$, the model predicts a saturation. For a non-conductive band tail ($E_\mathrm{\mu}=E_\mathrm{c}$), the simulations in Fig.\ref{fig:3}(a) showed that $E_\mathrm{F}$ lies in the band. If $E_\mathrm{F}$ lies in the band, the traps are filled and only $n_{\mathrm{mobile}}$ can still change with $E_{\mathrm{Fc}}$, therefore, when taking the derivative $\mathrm{d}V^\prime_{\mathrm{GB}}/\mathrm{d}E_{\mathrm{Fc}}$, we can set $n_{\mathrm{total}}=n_\mathrm{mobile}$ in (\ref{eq:vgb}). Further, for $u\gg 0$, $n_\mathrm{mobile}\!=\!g_\mathrm{c}\cdot k_\mathrm{B}T\cdot\ln(1+e^u)$ gives $n_{\mathrm{mobile}}=g_\mathrm{c}\cdot E_{\mathrm{Fc}}$. Differentiating $n_\mathrm{mobile}$ and $V_{\mathrm{GB}}^\prime$ to $E_\mathrm{Fc}$, combining these in (\ref{eq:ss}), and including the implicit $m$, yields
\begin{equation}
	SS=m^{\prime\prime}\cdot \frac{k_\mathrm{B}T_{c}^{\prime\prime}}{q}\cdot\ln(10),
	\label{eq:sslimit} 
\end{equation}
where $m^{\prime\prime}=m+q^2\cdot g_\mathrm{c}/C_{\mathrm{ox}}$ and $T_{\mathrm{c}}^{\prime\prime}=E_\mathrm{Fc0}/k_\mathrm{B}$. From numerical simulations we obtain that $E_\mathrm{Fc0}\!=\!\SI{0.15}{\micro\electronvolt}$, which gives $T_\mathrm{c}^{\prime\prime}=\SI{1.7}{\milli\kelvin}$, as shown in Fig.\ref{fig:oka}. Using Oka's experimental $C_{\mathrm{ox}}=\SI{5}{\milli\farad\per\meter\squared}$ (and $m^*$ and $m$ same as before) yields $SS=\SI{0.05}{\milli\volt}/\mathrm{dec}$, as shown in Fig.\ref{fig:oka}. From (\ref{eq:ids}), an expression for $E_\mathrm{Fc0}$ can be obtained, which gives 
\begin{equation}
	T_{\mathrm{c}}^{\prime\prime}=\frac{I_{\mathrm{DS}}\cdot L}{W\cdot \mu_0\cdot q\cdot g_\mathrm{c}\cdot V_{\mathrm{DS}}\cdot k_\mathrm{B}},
	\label{eq:Tcpp}
\end{equation}
which returns approximately the same $T_\mathrm{c}^{\prime\prime}$ ($\approx\!\SI{1.6}{\milli\kelvin}$) [here we used $I_\mathrm{DS}=\SI{e-10}{\ampere}$, $V_{\mathrm{DS}}\!=\!\SI{5}{\milli\volt}$, $W/L\!=\!1$, $\mu_0=\SI{2000}{\centi\meter\squared\per\volt\per\second}$, $m^*=1.08\cdot m_\mathrm{e}$, and $g_\mathrm{c}=m^*/(\pi\hbar^2)=\SI{4.51e14}{\per\centi\meter\squared\per\electronvolt}$]. This $T_\mathrm{c}^{\prime\prime}$ is below the base temperature of typical commercial dilution refrigerators ($\approx\SI{10}{\milli\kelvin}$). The temperature must drop below $T_{\mathrm{c}}^{\prime\prime}$ to test the hypothesis (\ref{eq:sslimit}) experimentally. However, (\ref{eq:Tcpp}) shows that $T_\mathrm{c}^{\prime\prime}$ can be increased by experimenting on lower mobility samples. 

\section{Conclusion}
From the presented theoretical analysis and simulations it can be concluded that $SS(T)$ saturates at sufficiently low temperatures. The nature of the band-tail states (conductive, traps, hybrid, or none) determines the plateau. Besides the standard plateau $m\cdot k_\mathrm{B}T_\mathrm{c}/q\cdot\ln(10)$ (conductive), two new plateau types were derived, $m^\prime \cdot k_\mathrm{B}T_\mathrm{c}^\prime/q\cdot\ln(10)$ (hybrid) and $m^{\prime\prime}\cdot k_\mathrm{B}T_\mathrm{c}^{\prime\prime}/q\cdot\ln(10)$ (traps, and none). The $T_\mathrm{c}^{\prime\prime}$ plateau is a testable hypothesis from the theory. With the $T$-axis in linear scale above 1\,K, the simulations showed that hybrid band tails can also explain a sudden drop of the standard plateau as well as an almost straight trend close to the thermal limit.

\bibliographystyle{ieeetr}
\bibliography{references}

\begin{thebibliography}{10}

\bibitem{tewksbury}
S.~Tewksbury, ``Attojoule {MOSFET} logic devices using low voltage swings and
  low temperature,'' {\em Solid-State Electronics}, vol.~28, pp.~255--276, Mar.
  1985.
\newblock
  \href{https://linkinghub.elsevier.com/retrieve/pii/0038110185900061}{doi:
  10.1016/0038-1101(85)90006-1}.

\bibitem{bohus}
H.~Bohuslavskyi, A.~G.~M. Jansen, S.~Barraud, V.~Barral, M.~Cass\'{e},
  L.~Le~Guevel, X.~Jehl, L.~Hutin, B.~Bertrand, G.~Billiot, G.~Pillonnet,
  F.~Arnaud, P.~Galy, S.~De~Franceschi, M.~Vinet, and M.~Sanquer, ``Cryogenic
  {Subthreshold} {Swing} {Saturation} in {FD}-{SOI} {MOSFETs} {Described}
  {With} {Band} {Broadening},'' {\em IEEE Electron Device Letters}, vol.~40,
  pp.~784--787, May 2019.
\newblock \href{https://ieeexplore.ieee.org/document/8660508/}{doi:
  10.1109/LED.2019.2903111}.

\bibitem{edl}
A.~Beckers, F.~Jazaeri, and C.~Enz, ``Theoretical {Limit} of {Low}
  {Temperature} {Subthreshold} {Swing} in {Field}-{Effect} {Transistors},''
  {\em IEEE Electron Device Letters}, vol.~41, pp.~276--279, Feb. 2020.
\newblock \href{https://ieeexplore.ieee.org/document/8946710/}{doi:
  10.1109/LED.2019.2963379}.

\bibitem{ghibaudo}
G.~Ghibaudo, M.~Aouad, M.~Cass\'{e}, S.~Martinie, T.~Poiroux, and F.~Balestra,
  ``On the modelling of temperature dependence of subthreshold swing in
  {MOSFETs} down to cryogenic temperature,'' {\em Solid-State Electronics},
  vol.~170, p.~107820, Aug. 2020.
\newblock
  \href{https://linkinghub.elsevier.com/retrieve/pii/S0038110120300812}{doi:
  10.1016/j.sse.2020.107820}.

\bibitem{tnano}
A.~Beckers, J.~Michl, A.~Grill, B.~Kaczer, M.~G. Bardon, B.~Parvais,
  B.~Govoreanu, K.~De~Greve, G.~Hiblot, and G.~Hellings, ``{Physics-Based and
  Closed-Form Model for Cryo-CMOS Subthreshold Swing},'' {\em IEEE Transactions
  on Nanotechnology}, vol.~22, pp.~590--596, 2023.
\newblock \href{https://ieeexplore.ieee.org/abstract/document/10250972}{doi:
  10.1109/TNANO.2023.3314811}.

\bibitem{ghibaudo2}
G.~Ghibaudo, M.~Aouad, M.~Cass\'{e}, T.~Poiroux, and C.~Theodorou, ``{On the
  diffusion current in a MOSFET operated down to deep cryogenic
  temperatures},'' {\em Solid-State Electronics}, vol.~176, p.~107949, 2021.
\newblock
  \href{https://www.sciencedirect.com/science/article/pii/S0038110120304160}{doi:
  10.1016/j.sse.2020.107949}.

\bibitem{hafez}
I.~M. Hafez, G.~Ghibaudo, and F.~Balestra, ``Assessment of interface state
  density in silicon metal-oxide-semiconductor transistors at room,
  liquid-nitrogen, and liquid-helium temperatures,'' {\em Journal of Applied
  Physics}, vol.~67, pp.~1950--1952, Feb. 1990.
\newblock \href{http://aip.scitation.org/doi/10.1063/1.345572}{doi:
  10.1063/1.345572}.

\bibitem{jock}
R.~M. Jock, S.~Shankar, A.~M. Tyryshkin, J.~He, K.~Eng, K.~D. Childs, L.~A.
  Tracy, M.~P. Lilly, M.~S. Carroll, and S.~A. Lyon, ``Probing band-tail states
  in silicon metal-oxide-semiconductor heterostructures with electron spin
  resonance,'' {\em Applied Physics Letters}, vol.~100, p.~023503, Jan. 2012.
\newblock \href{http://aip.scitation.org/doi/10.1063/1.3675862}{doi:
  10.1063/1.3675862}.

\bibitem{oka}
H.~Oka, H.~Asai, T.~Inaba, S.~Shitakata, H.~Yui, H.~Fuketa, S.~Iizuka, K.~Kato,
  T.~Nakayama, and T.~Mori, ``Milli-{Kelvin} {Analysis} {Revealing} the {Role}
  of {Band}-edge {States} in {Cryogenic} {MOSFETs},'' in {\em 2023
  {International} {Electron} {Devices} {Meeting} ({IEDM})}, (San Francisco, CA,
  USA), pp.~1--4, IEEE, Dec. 2023.
\newblock \href{https://ieeexplore.ieee.org/document/10413872/}{doi:
  10.1109/IEDM45741.2023.10413872}.

\bibitem{yurtta}
N.~Yurttagül, M.~Kainlauri, J.~Toivonen, S.~Khadka, A.~Kanniainen, A.~Kumar,
  D.~Subero, J.~Muhonen, M.~Prunnila, and J.~S. Lehtinen, ``Millikelvin
  {Si}-{MOSFETs} for {Quantum} {Electronics},'' Oct. 2024.
\newblock \href{http://arxiv.org/abs/2410.01077}{arXiv:2410.01077 [cond-mat,
  physics:physics]}.

\bibitem{gen}
A.~Beckers, D.~Beckers, F.~Jazaeri, B.~Parvais, and C.~Enz, ``{Generalized
  Boltzmann relations in semiconductors including band tails},'' {\em Journal
  of Applied Physics}, vol.~129, p.~045701, Jan. 2021.
\newblock
  \href{https://pubs.aip.org/aip/jap/article-pdf/doi/10.1063/5.0037432/14773660/045701_1_online.pdf}{doi:
  10.1063/5.0037432} Available: \href{https://arxiv.org/abs/2309.13687}{arXiv:
  2309.13687}.

\end{thebibliography}

\end{document}